\documentclass[11pt]{article}
\usepackage[pdftex]{graphicx}       
\begin{document}

\begin{center}{\Large Revisiting the Two-Filter Formula for Smoothing for State-Space Models.}\\

\vspace{10mm}
{\large Genshiro Kitagawa}\\[2mm]
The Institute of Statistical Mathematics\\
and\\
Graduate University for Advanced Study

\vspace{2mm}
{\today}
\end{center}

\noindent{\bf Abstract}

\noindent
Smoothing algorithms for state-space models, i.e., fixed-interval smoothing, fixed-lag smoothing, and two-filter formula for smoothing, are examined using real examples. 
For linear and Gaussian state-space models, it is observed that similar posterior distributions can be obtained by properly defining the inverse filter.
In the case of linear non-Gaussian state-space models, it is shown that Gaussian-sum smoothing is possible even for relatively high dimensional state-space model with Gaussian-mixture noise inputs by properly setting the inverse filter.
The two-filter formula is also applicable for particle filter, but better results are obtained with fixed lag smoothing or with the average of forward and backward fixed lag smoothers.

\vspace{1mm}
\noindent{\bf Key words and phrases:} 

\noindent
Non-Gaussian smoother, Gaussian-sum smother, particle smoother, Gaussian mixture noise, nonstationary time series, outliers, seasonal adjustment.

\section{Introduction}

Non-Gaussian state-space modeling is now an important analytical tool in time series analysis, particularly useful for analyzing time series with abrupt changes in structure or outliers, and for analyzing discrete or nonlinear processes. See, for example, West and Harrison (1989), Doucet et al. (2001) and references therein.

Kitagawa (1987) presented an implementation of a non-Gaussian smoothing algorithm based on a numerical approximation of the associated probability distribution.
Although this method has a wide variety of applications, its application to problems with high state dimensionality (e.g., more than 4 dimensions) is impractical because it requires computationally expensive numerical integration.
In actual time series analysis, many problems require higher-dimensional state vectors.
For example, seasonal adjustment of monthly time series requires a state vector of at least 13 dimensions (Kitagawa and Gersch (1984), Kitagawa (1989)).
Kitagawa (1989) modeled seasonal time series using a state-space model in which a mixed Gaussian distribution is assumed to be the system noise or observation noise in order to handle abrupt changes in trend and seasonal components and outliers in seasonal data. Modeling using mixed Gaussian distributions has been proposed by Sorenson and Alspach (1971), Alspach and Sorenson (1972), Harrison and Stevens (1976), and Anderson and Moore (1979), but smoothing does not address smoothing.
However, smoothing problems are very important in statistical data analysis. Kitagawa (1989) realized fixed-lag smoothing of a non-Gaussian state-space model using a high-dimensional state vector (about 40 dimensions) and achieved a significant improvement in estimates of seaso

Based on this experience, we were motivated to revisit the implementation of practical smoothing algorithms for general state-space models.
In this paper, we first review filtering, fixed-interval smoothing, and smoothing algorithms with two-filter formulas for general state-space models.
Next, we revisit the smoothing algorithms for three cases: a standard linear Gaussian state-space model, a linear state-space model with a mixed Gaussian noise distribution, and a general state-space model.
Specifically, we revisit the smoothing algorithms for the Kalman filter, Gaussian-sum filter, and particle filter for the three models, using the time series used in previous analyses.

The plan of this paper is as follows. In Section 2, we briefly present the recursive filtering and smoothing algorithm and the two-filter formula for smoothing for the fixed-interval smoothing algorithm.
In Section 3, we examine the two-filter smoothing algorithm in the case of the standard state-space model and show the results with seasonal time series.
In Section 4, we examine the Gaussian sum filter and smoothing algorithm in the case of a mixed Gaussian distribution of noise, and compare the results using artificial time series data with jumps in the trend.
Section 5 examines the smoothing algorithm for the particle filter and presents results from trend estimation and seasonal adjustment. Section 6 provides a summary of the entire report.

\section{A Brief Review of the Filtering and Smoothing Algorithms}
\subsection{The state-space model and the state estimation problems}

Assume that a time series $y_n$ is expressed by a nonlinear state-space model
\begin{eqnarray}
  x_n &=& f(x_{n-1}) + g(v_n) \nonumber \\
  y_n &=& h(x_n) + w_n, \label{Eq_GSSM}
\end{eqnarray}
where $f$, $g$ and $h$ are possibly nonlinear function, $x_n$ is an $d_x$-dimensional state vector, $v_n$ and $w_n$ are $d_v$-dimensional and
$\ell$-dimensional white noise sequences having density functions $q_n(v)$ and $r_n(w)$,
respectively. 
The initial state vector $x_0$ is assumed to be distributed according to
the density $p(x_0)$. 
The observations from time $i$ to $j$, $\{y_i,\ldots ,y_j\}$, is denoted by $Y_{i:j}$.
For simlicity, $Y_{1:j}$ is denoted as $Y_j$.
The problem of state estimation is to evaluate
$p(x_n |Y_j)$, the conditional density of $x_n$ given the observations $Y_j$ 
and the initial density $p(x_0|Y_0) = p(x_0)$. 
For $n> j$, $n= j$ and $n< j$, the problems are referred to as  
prediction, filtering and smoothing, respectively.

\subsection{The generic filter and smoother}

In Kitagawa (1987), it was shown that for the state-space model (\ref{Eq_GSSM}) with non-Gaussian 
white noise $v_n$ and $w_n$, the recursive formulas for obtaining the densities
of the one step ahead predictor, the filter and the smoother are as follows:

\vspace{2mm}
One step ahead prediction:
\begin{eqnarray}
p(x_n |Y_{n-1}) = \int_{-\infty}^{\infty}  p(x_n |x_{n-1})p(x_{n-1}|Y_{n-1})dx_{n-1}.
\end{eqnarray}

\vspace{2mm}
Filtering:
\begin{eqnarray}
p(x_n|Y_n)= \frac{ p(y_n |x_n )p(x_n |Y_{n-1})}{p(y_n |Y_{n-1})},
\end{eqnarray}
where $p(y_n|Y_{n-1})$ is obtained by $\displaystyle \int_{-\infty}^{\infty} \! p( y_n|x_n )p(x_n |Y_{n-1})dx_n$.

\vspace{2mm}
Fixed-interval smoothing: For $n=N-1,\ldots ,1$,
\begin{eqnarray}
p(x_n|Y_N ) = p (x_n|Y_n ) \int_{-\infty}^{\infty} \frac{p(x_{n+1}|Y_N )p(x_{n+1}|x_n )}
{p(x_{n+1}|Y_n )} dx_{n+1}.
\end{eqnarray}

\subsection{The Two-filter Formula for Smoothing}

In this subsection, two-filter formula for smoothing is briefly introduced 
which is an alternative algorithm of fixed-interval smoothing (Fraser (1967), Mayne (1966), Kitagawa (1994)).
Firstly, since $Y_N = Y_{n-1}\cup Y_{n:N}$, the smoothed density $p(x_n|Y_N )$ can be expressed as
follows:
\begin{eqnarray}
p(x_n|Y_N) &=& p(x_n |Y_{n-1}, Y_{n:N}) \nonumber \\
   &=& p(x_n, Y_{n -1}, Y_{n:N})p(Y_{n-1}, Y_{n:N})^{-1} \nonumber \\
   &=& p(x_n, Y_{n -1}, Y_{n:N})p(Y_{n-1})^{-1}p(Y_{n:N}|Y_{n-1})^{-1} \nonumber \\
   &=& p(x_n, Y_{n:N} |Y_{n -1})p(Y_{n:N} |Y_{n-1})^{-1} \nonumber \\
   &=& p(x_n |Y_{n-1})p(Y_{n:N} |x_n , Y_{n-1})p(Y^n |Y_{n -1})^{-1} \nonumber \\
   &=& p(x_n |Y_{n-1})p(Y_{n:N} |x_n )p(Y_{n:N} |Y_{n-1})^{-1}. \label{Eq_two-filter-formula1}
\end{eqnarray}
Note that we can similarly obtain two-filter formula based on the filter distribution:
\begin{eqnarray}
p(x_n|Y_N) = p(x_n |Y_n)p(Y_{n+1:N} |x_n )p(Y_{n+1:N} |Y_{n})^{-1}. 
\end{eqnarray}
Since $p(x_n |Y_{n-1})$ and $p(x_n |Y_n)$ have been already given by forward filtering,
 and $p(Y_{n:N} |Y_{n-1})$ and $p(Y_{n+1:N} |Y_n)$
are constants which do not depend on $x_n$, the smoothed density $p(x_n|Y_N)$ can be obtained
if $p(Y_{n:N} |x_n )$ or $p(Y_{n+1:N} |x_n )$ is given. Here these terms can be 
evaluated by the following backward filtering:

\vskip2mm

Initialization
\begin{eqnarray}
  p(Y_{N:N}|x_N) = p(y_N|x_N).
\end{eqnarray}

Backward prediction
\begin{eqnarray}
  p(Y_{n+1:N}|x_n) &=& \int_{-\infty}^{\infty} p(Y_{n+1:N},x_{n+1}|x_n) dx_{n+1} \nonumber \\
   &=& \int_{-\infty}^{\infty} p(Y_{n+1:N},x_{n+1},x_n) p(x_{n+1}|x_n) dx_{n+1}  \\
   &=& \int_{-\infty}^{\infty} p(Y_{n+1:N}|x_{n+1}) p(x_{n+1}|x_n) dx_{n+1}. \nonumber 
\end{eqnarray}

Backward filtering
\begin{eqnarray}
  p(Y_{n:N}|x_n) &=& p(y_n,Y_{n+1:N}|x_n)  \nonumber \\
   &=&  p(y_n|x_n,Y_{n+1:N})p(Y_{n+1:N}|x_n) \nonumber  \\
   &=&  p(y_n|x_n)p(Y_{n+1:N}|x_n).    
\end{eqnarray}

\section{Linear-Gaussian Case}

\subsection{The Kalman filter and the smoother}

In this section, we assume that the state-space model is linear and is given by
\begin{eqnarray}
  x_n &=& F_n x_{n-1} + G_n v_n \nonumber \\
  y_n &=& H_n x_n + w_n,
\end{eqnarray}
where $F$, $G$ and $H$ are $d_x \times d_x$, $d_x \times d_v$ and $1 \times d_x$ dimensional matrices, respectively.  
It is well known that if all of the noise densities $q_n(v)$ and $r_n(w)$ and the initial
state density $p(x_0)$ are Gaussian, then the conditional density $p(x_n|Y_m)$ is also
Gaussian and that the mean and the variance-covariance matrix can be obtained by the Kalman filter and the fixed interval smoothing algorithms (Anderson and Moore (1979)).

To be specific, if we assume $q_n(v) \sim N(0, Q_n)$, $r_n(w)\sim N(0, R_n)$, $p(x_0|Y_0 ) \sim
N(x_{0|0}, V_{0|0})$ and $p(x_n|Y_m ) \sim N( X_{n|m}, V_{n|m})$, then the Kalman filter consists of the following sequential computations for $n=1,\ldots ,N$:

\vskip2mm
\textbf{One-step ahead prediction}
\begin{eqnarray}
x_{n|n-1} &=& F_n x_{n-1|n-1}, \nonumber \\
V_{n|n-l} &=& F_n V_{n-1|n-1} F_n^T + G_nQ_nG_n^T.
\end{eqnarray}

\textbf{Filter}
\begin{eqnarray}
K_n  &=& V_{n|n-1}H_n^T(H_nV_{n|n-1}H_n^T + R_n)^{-1}, \nonumber \\
x_{n|n} &=& x_{n|n-1} + K_n(y_n - H_n x_{n|n-1}), \\
V_{n|n} &=& (I-K_n H_n)V_{n|n-1}. \nonumber
\end{eqnarray}
Using these estimates, the smoothed density of the state $x_n$ given the data $Y_N$ is obtained by the following backward recursion for $n=N-1,\ldots ,1$:

\vskip2mm
\textbf{Fixed interval smoothing algorithm}
\begin{eqnarray}
A_n     &=& V_{n|n} F_n^T V_{n+1|n}^{-1}, \nonumber \\
x_{n|N} &=& x_{n|n} + A_n(x_{n+1|N} - x_{n+1|n}), \\
V_{n|N} &=& V_{n|n} + A_n (V_{n+1|N} - V_{n+1|n})A_n^T. \nonumber
\end{eqnarray}
Note that the initial values for this recursion, $x_{N|N}$ and $V_{N|N}$, are obtained by the Kalman filter.

\subsection{Two-filter formula for the linear Gaussian state-space model}

The two-filter formula for the linear Gaussian state-space model is considered here.
Note that the fixed-interval smoothing algorithm is applicable to the linear Gaussian state-space model.
Therefore, the use of the two-filter formula is not essential, and the purpose here is to explain the backward filter, compare it with fixed-interval smoothing, and prepare its application to Gaussian sum smoothing and particle smoothing.

In the case of the linear Gaussian state-space model, from equation (\ref{Eq_two-filter-formula1}),
 the contitional density of the 
state $x_n$ given the observations $Y_N$ is Gaussian and can be expressed as follows:
\begin{eqnarray}
  p(x_n|Y_N) = \varphi (Y_{n:N}|x_n) \varphi (x_n|Y_{n-1}) = \varphi (x_n|Y_N) ,
\end{eqnarray}
where $\varphi$ denotes a Gaussian density function.
If we assume that $\varphi (x_n|Y_{n-1})\sim N(x_{n|n-1},V_{n|n-1})$ and 
$\varphi (x_n|Y_{n:N})\sim N(z_{n|n},U_{n|n})$, then the smoothed density of the state is also Gaussian, $\varphi (x_n|Y_N)\sim N(x_{n|N},V_{n|N})$, and its mean and the variance-covariance matrix are obtained by
\begin{eqnarray}
   J_n     &=& V_{n|n-1}\left(V_{n|n-1} + U_{n|n}\right)^{-1} \nonumber \\
   x_{n|N} &=& x_{n|n-1} + J_n (z_{n|n} - x_{n|n-1}) \\
   V_{n|N} &=& (I - J_n) V_{n|n-1}.  \nonumber
\end{eqnarray}

The details of the backward filter is shown in Kitagawa (2023).
Given the mean and the varaiance-covariance matrix at the end point,  $z_{N|N}$ and $U_{N|N}$, 
$z_{n|n}$ and $U_{n|n}$ can be obtained by using the backward (reverse) state-space model:
\begin{eqnarray}
  z_n &=& F_{n+1}^{-1}z_{n+1} - F_{n+1}^{-1}G_{n+1}v_{n+1}^{(B)} \nonumber \\
  y_n &=& H_n z_n + w_n^{(B)},  \nonumber
\end{eqnarray}
where $v_{n+1}^{(B)} \sim N(0,Q_{n+1})$ and $w_n^{(B)} \sim N(0,R_n)$.
Therefore, we can apply the same Kalman filter by replacing $F_n$ and $G_n$ by
$\bar{F}_{n+1} = F_{n+1}^{-1}$ and $\bar{G}_{n+1} = -F_{n+1}^{-1}G_{n+1}$. 
Namely, for $n=N-1,\ldots ,1$, the backward filter is given by:

\vskip2mm
\textbf{Backward one-step ahead prediction:}
\begin{eqnarray}
z_{n|n+1} &=& \bar{F}_{n+1} z_{n+1|n+1}, \nonumber \\
U_{n|n+l} &=& \bar{F}_{n+1} U_{n+1|n+1} \bar{F}_{n+1}^T + \bar{G}_{n+1}Q_{n+1}\bar{G}_{n+1}^T.\nonumber
\end{eqnarray}

\textbf{Backward filter:}
\begin{eqnarray}
\bar{K}_n  &=& U_{n|n+1}H_n^T(H_nU_{n|n+1}H_n^T + R_n)^{-1} \nonumber \\
z_{n|n} &=& z_{n|n+1} + \bar{K}_n(y_n - H_n z_{n|n+1}), \\
U_{n|n} &=& (I- \bar{K}_n H_n)U_{n|n+1}. \nonumber
\end{eqnarray}

A practical way to define the initial distribution is to put
\begin{eqnarray}
 z_{N|N} = x_{N|N},\quad U_{N|N} = V_{N|N}.
\end{eqnarray}
This allows the inverse filter to be started, but as will be seen in later example, the smoothed distribution obtained in this way is evaluated with a smaller error variance than the exact smoothed distribution in the first part.
One way to improve this problem is to increase the diagonal elements of the initial variance-covariance matrix as $\nu^2/m$, where $\nu$ is the variance of the time series in one cycle, i.e., $\nu^2 = p^{-1}\sum_{j=1}^p (y_{N-j+1}-\mu)$, $\mu = p^{-1}\sum_{j=1}^p y_{N-j+1}$, $p$ is the length of one cycle and $m$ is the dimension of the state. 
Kitagawa (1994) show a method of defining the initial state vector to garantee the full-rank of the initial variance-covariance matrix.

On the other hand, Balenzuela et al. (2022) proposed an exact smoothing algorithm based on the backward information filter. 
Hereafter, we shall briefly show a method based on the information filter (Kaminski 1971) for the backward state-space model. The details of this method and the relationship with the Bakenzuela's algorithm is shown in Kitagawa (2023).
The information filter computes sequentially not the variance-covariance matrix $U_{n|n}$ but its inverse (information matrix) $U_{n|n}^{-1}$.
Further instead of the state vector, $d_{n|n-1} = U_{n|n+1}^{-1} z_{n|n+1}$ and $d_{n|n} = U_{n|n}^{-1} z_{n|n}$ are updated.
Applying the information filter to the backward state-space model, we obtain the following backward information filter:

\vspace{5mm}
\textbf{Information predictor}
\begin{eqnarray}
   L_n  &=& -F_{n+1}^{T} U_{n-1|n-1}^{-1} G_{n+1} (Q_{n+1}^{-1} + G_{n+1}^TU_{n+1|n+1}^{-1}G_{n+1})^{-1}, \nonumber \\
  d_{n|n+1} &=& (F_{n+1}^T +L_nG_{n+1}^T) d_{n+1|n+1}, \nonumber \\
  U_{n|n+1}^{-1} &=& (F_{n+1}^T + L_nG_{n+1}^T) U_{n+1|n+1}^{-1} F_{n+1}. \nonumber  
\end{eqnarray}

\textbf{Information filter}
\begin{eqnarray}
  d_{n|n} &=& d_{n|n+1} + H_n^T R_n^{-1} y_n,  \nonumber \\
  U_{n|n}^{-1} &=& U_{n|n+1}^{-1} + H_n^T R_n^{-1} H_n. \nonumber 
\end{eqnarray}

Then the smoothed distribution of the state, $x_{n|N}$ and $V_{n|N}$, are obtained by:
\textbf{Two-Filter Formula for Smoothing}
\begin{eqnarray}
  J_n     &=& (V_{n|n}^{-1} + U_{n|n+1}^{-1})^{-1}U_{n|n+1}^{-1} \nonumber \\
  V_{n|N} &=& (V_{n|n}^{-1} + U_{n|n+1}^{-1})^{-1} \label{Eq_Two-filter_inofrmation-matrix_version} \\
  x_{n|N} &=& V_{n|N}(V_{n|n}^{-1}x_{n|n} + d_{n|n+1}). \nonumber
\end{eqnarray}

The derivation of these recursive formula is shown in Kitagawa (2023).
Note that in the case of backward information filter, the inverse of the
transition matrix $F_n^{-1}$ is replaced by the transpose of the matrix
$F_{n+1}^T$ and that the two-filter formula of equation (\ref{Eq_Two-filter_inofrmation-matrix_version}) does not require the matrix inversion of $U_{n|n+1}^{-1}$.


\subsection{Example: Seasonal adjustment model}
For the explanation of the two filter formula, we re-analize the BLSALLFOOD data analysed in 
Kitagawa and Gersch (1974) and Kitagawa (2020)). We consider the seasonal adjustment models 
with stationary AR component:
\begin{eqnarray}
   y_n = T_n + S_n + p_n + w_n,
\end{eqnarray}
where $T_n$, $S_n$ and $p_n$ are the trend, the seasonal component and the stationary AR component 
that follow the following component models,
\begin{eqnarray}
   T_n &=& 2T_{n-1} - T_{n-2} + u_n \nonumber \\
   S_n &=& -(S_{n-1}+\cdots + S_{n-11}) + v_n \\
   p_n &=& \sum_{j=1}^{m_3} a_j p_{n-j} + z_n, \nonumber
\end{eqnarray}
where $m_3$ is the order of AR model, $u_n \sim N(0,\tau^2_1)$, $v_n \sim N(0,\tau^2_2)$, $z_n \sim N(0,\tau^2_3)$ and $w_n \sim N(0,\sigma^2)$.

\begin{figure}[tbp]
\begin{center}
\includegraphics[width=160mm,angle=0,clip=]{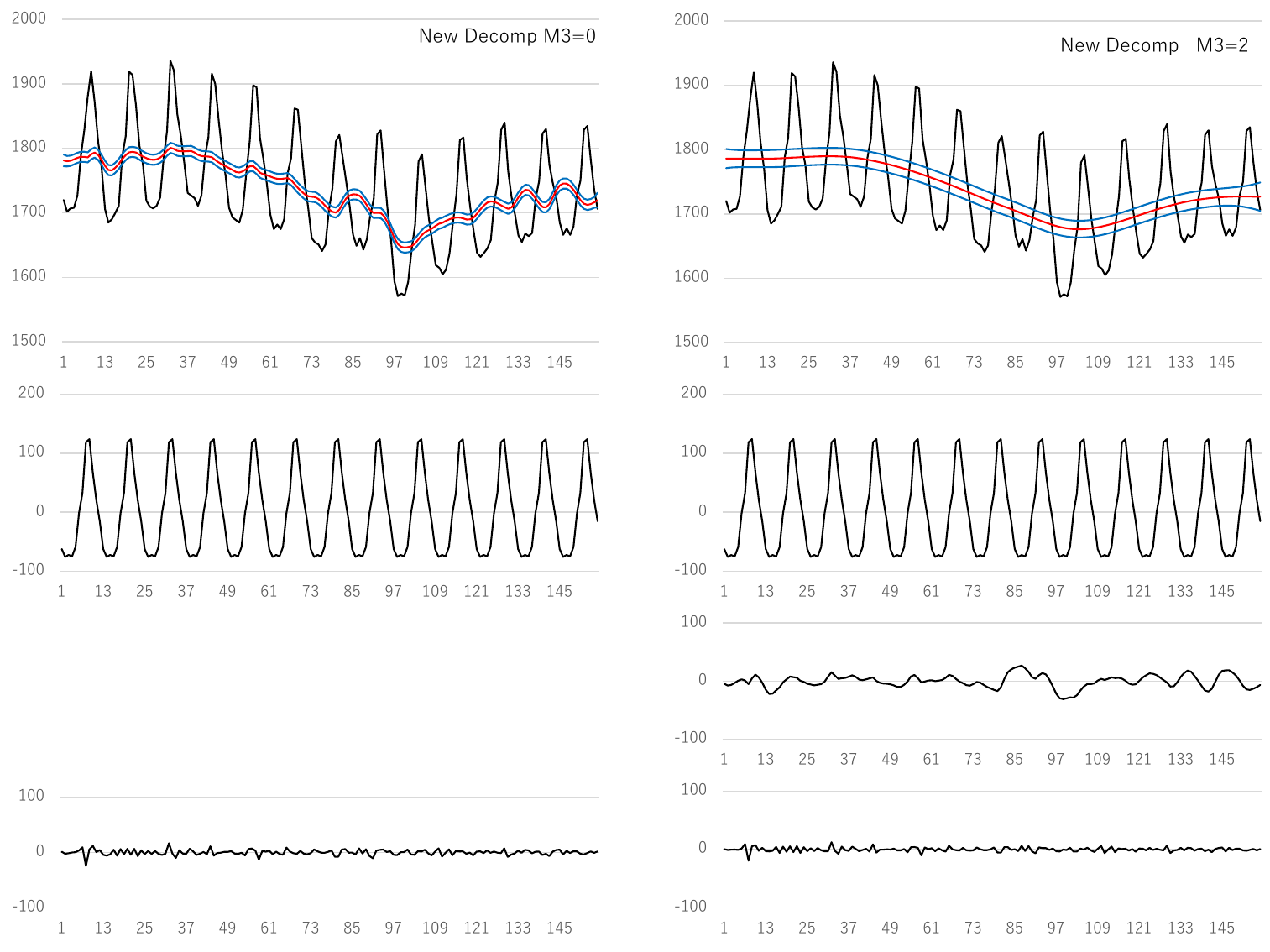}
\end{center}
\caption{The seasonal adjustment by the fixed-interval smoothing algorithm.
Left plots: without AR component $m_3=0$, Right plots: with AR component, $m_3=2$. 
Top plots show the data (black) and the mean (red) and $\pm$2 standard error (blue) of the trend, the second plots the seasonal component, the third plots the AR component and the bottom plots show the noise component. }
\label{Fig_Gauss_fixed-interval-smoothing}
\end{figure}

\begin{table}[bp]
\caption{Estimated parameters of two seasonal adjustment models}
\label{Tab_Estimated-parameters of TSAR}
\begin{center}
\begin{tabular}{c|c|c}
              & $m_3=0$                & $m_3=2$      \\
\hline
$\tau_1^2$ &     21.0870             & 0.17605 \\
$\tau_2^2$ & $0.37237\!\times\! 10^{-5}$ & $0.98741\!\times\! 10^{-3}$ \\
$\tau_3^2$ & \rule[1mm]{8mm}{0.05mm} &     29.616             \\
$\sigma^2$ &     37.274              &     29.616             \\
$a_1$      & \rule[1mm]{8mm}{0.05mm} &     1.30754           \\
$a_2$      & \rule[1mm]{8mm}{0.05mm} &    $-0.47758$         \\
\hline
\end{tabular} 
\end{center}
\end{table}
The maximum likelihood estimates of the parameters of the models for $m_3=0$ and 2 are 
shown in Table \ref{Tab_Estimated-parameters of TSAR}.

Figure \ref{Fig_Gauss_fixed-interval-smoothing} shows the decomposition of the seasonal data
by the fixed-interval smoothing algorithm based on the estimated models.
Left plots show the estimated trend, seasonal and noise components obtained
by the model with $m_3=0$.
In the top plot, the black curve shows the original data, the red one the mean of the
trend component which is given as the first component of $x_{n|N}$.
Two blue curves indicate $\pm$2 standard error interval of the trend.
Right plots show the estimates by the model with AR component with $m_3=2$, and the third plot from the top shows the estimated AR component. 
Very smooth trend was obtained by the seasonal adjustment model
with AR component.

\begin{figure}[tbp]
\begin{center}
\includegraphics[width=160mm,angle=0,clip=]{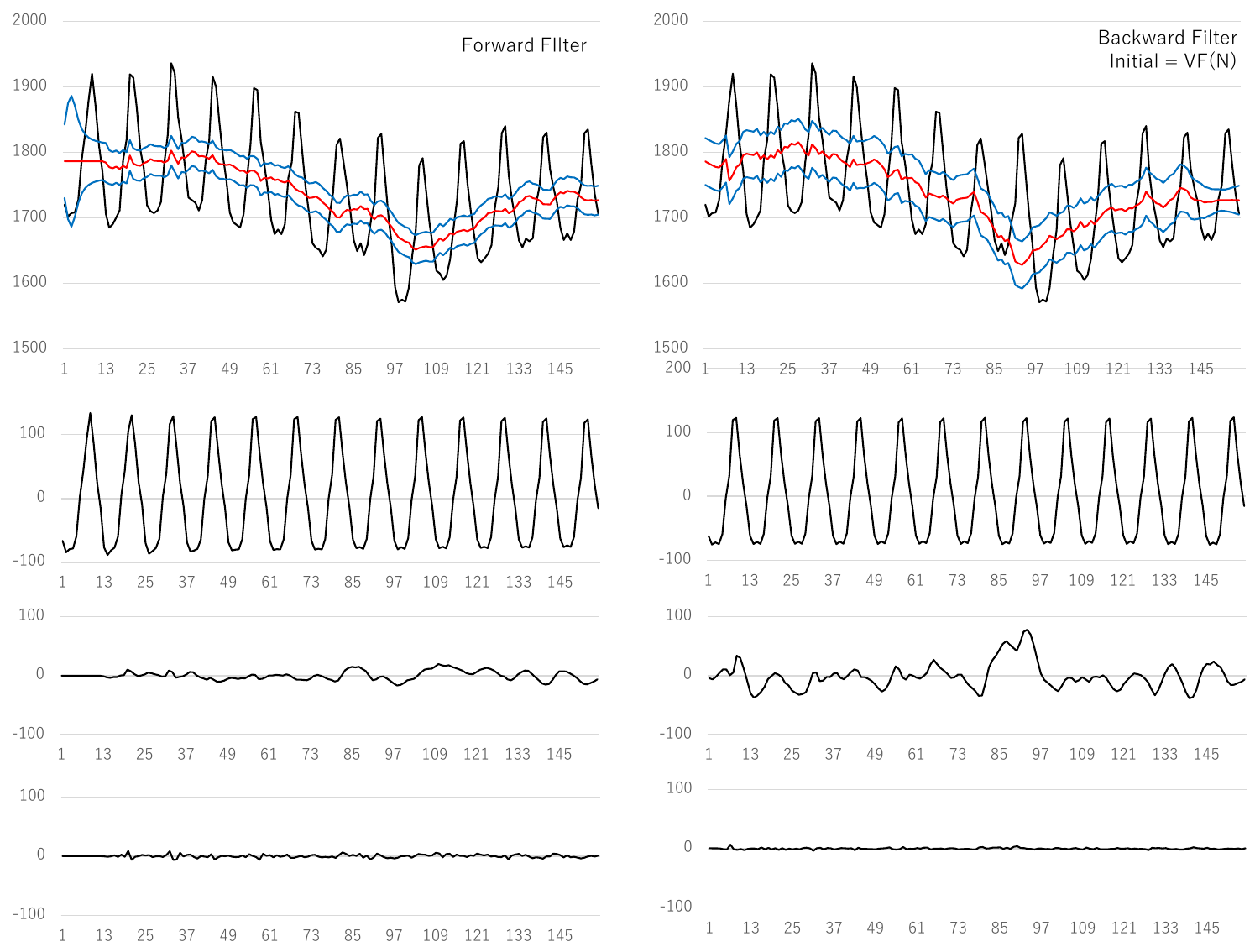}
\end{center}
\caption{Decomposition by the forward filter and the backward filter with initial condition
 $z_{N|N} = x_{N|N}$ and $U_{N|N} = V_{N|N}$.
Left plots: forward filter, Right plots: backward filter. 
Top plot shows the data (black), the mean of the trend (red), $\pm 2$ standard error (blue),
the second plot the seasonal component, the third plot the AR component and the bottom plot shows the noise component. }
\label{Fig_Gauss_forward_backward_filters}
\end{figure}

Figure \ref{Fig_Gauss_forward_backward_filters} shows the intermidiate results of the 
two-filter method for smoothing when we applied to the model with $m_3=2$.
The left plots show the decomposition by the forward filter,
which are obtained through $x_{n|n}$.
The right plots show the results by the backward filter when the filter estimates 
$x_{N|N}$ and $V_{N|N}$ are used as the initial distribution.
The $\pm 2$ standard error intervals for the backward filter are generally wider than those for the forward filter, but the first part of the backward filter is found out to be too narrow.
The AR component of the backward filter is more variable than the forward filter.
This is considered because the backward AR component model is nonstationary.

\begin{figure}[tbp]
\begin{center}
\includegraphics[width=160mm,angle=0,clip=]{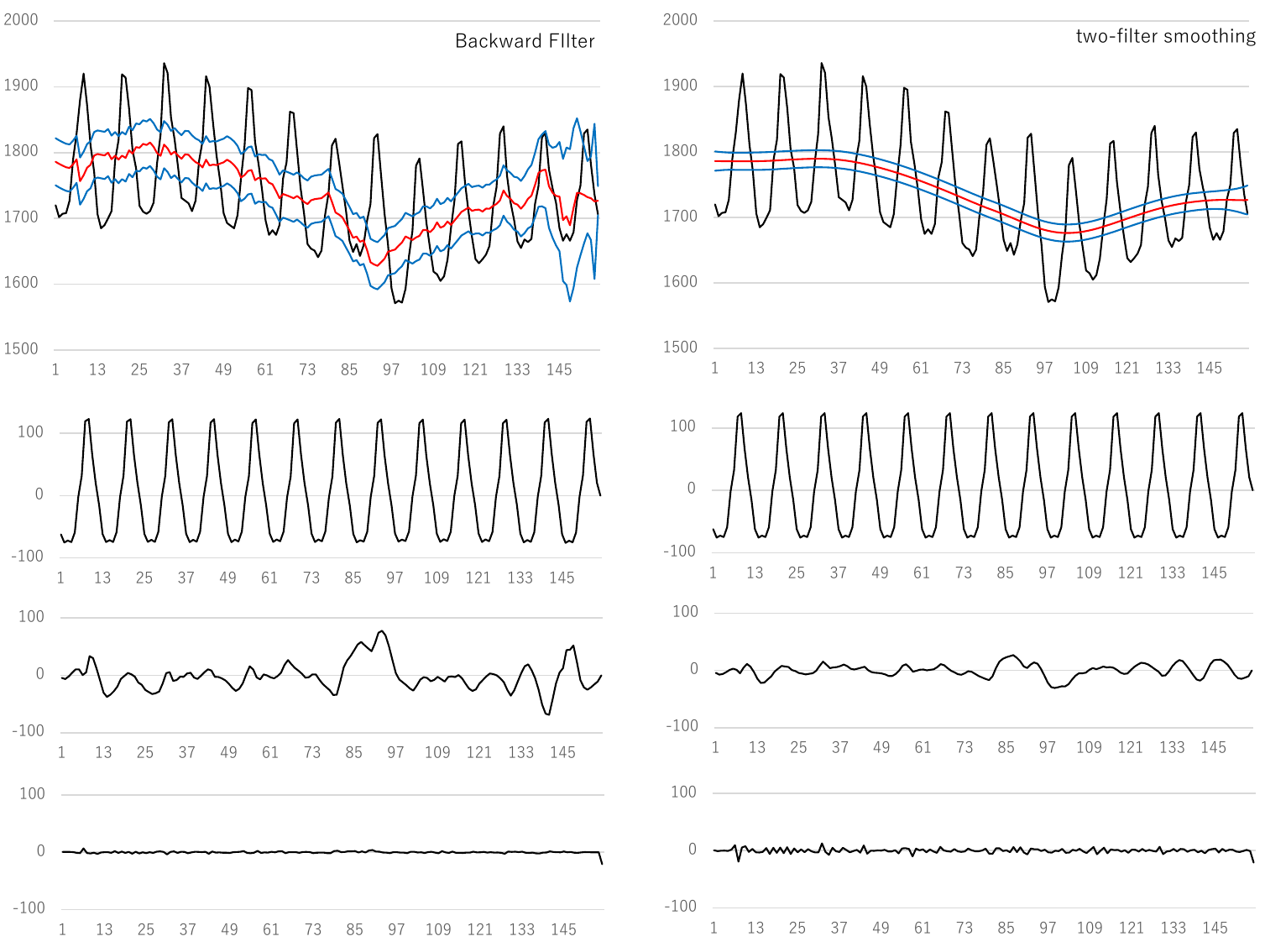}
\end{center}
\caption{The seasonal adjustment by the two-filter formula for smoothing.
Left plots: without AR component $m_3=0$, Right plots: with AR component, $m_3=2$. 
Top plot shows the data (black), the mean of the trend (red), $\pm 2$ standard error (blue), 
the second plot the seasonal component, the third plot the AR component and the bottom plot shows the noise component. }
\label{Fig_Gauss_two-filter-formula_1}
\end{figure}

Figure \ref{Fig_Gauss_two-filter-formula_1} shows the decomposition by the backward filter
with increased initial variance-covariance matrix $U_{N|N}$ and the two-filter formula for smoothing, respectively.
With the increased variances for the initial variance-covariance matrix, the standard error interval shown in the top left plot becomes significantly wider than the one in the top right plot of Figure \ref{Fig_Gauss_forward_backward_filters}.
The trend estimate by the two-filter smoother becomes very smooth and the $\pm$2 confidence interval becomes considerably narrower than both the forward predictor and the backward filter.
Compared with the right plots of Figure \ref{Fig_Gauss_fixed-interval-smoothing},
we can see that, at least visually, the estimates of the trend, seasonal component, 
AR component and the noise are indistinguishable 
with the fixed interval smoothing estimates.

\begin{figure}[tbp]
\begin{center}
\includegraphics[width=160mm,angle=0,clip=]{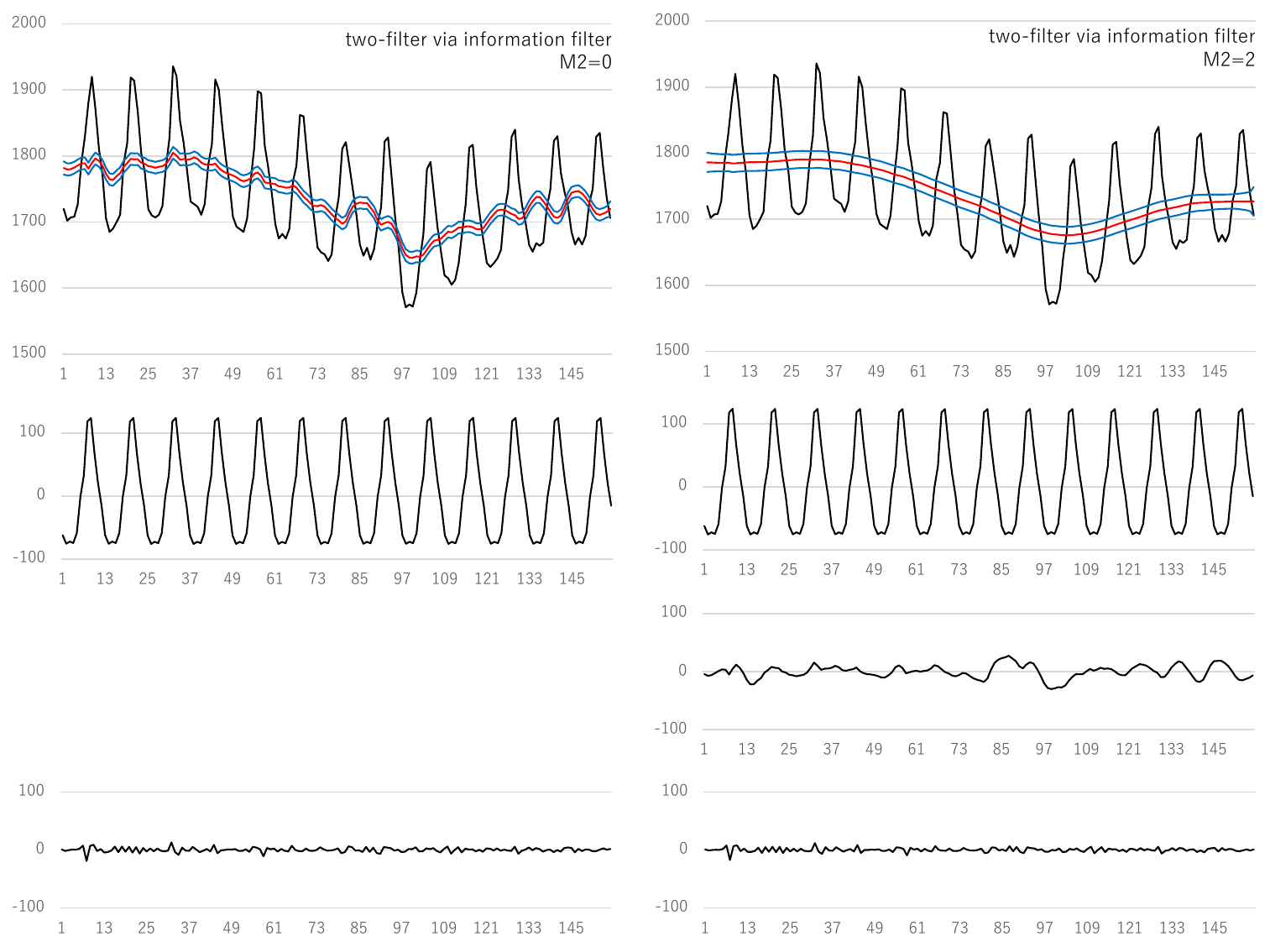}
\end{center}
\caption{The seasonal adjustment by the two-filter formula for smoothing.
Left plot: without AR component $m_3=0$, Right plot: with AR component, $m_3=2$. 
Top plot shows the data (black), the mean of the trend (red), $\pm 2$ standard error (blue), 
the second plot the seasonal component, the third plot the AR component and the bottom plot shows the noise component. }
\label{Fig_Gauss_two-filter-formula_2}
\end{figure}

Figure \ref{Fig_Gauss_two-filter-formula_2} shows the results of smoothing 
by the two-filter formula using the backward information filter shown in Section 3.2.
Left plots show the case of $m_3=0$, i.e., there are no AR component in the model.
The fiexed-interval smoothing estimates for this model is shown in the left plots of
Figure \ref{Fig_Gauss_fixed-interval-smoothing}.
The right plots show the estimates by the model with $m_3=2$.
The estimates obtained by this two-filter formula are generally consistent with those obtained by fixed-interval smoothing, but at the extreme end ($n>150$), the standard error interval is slightly narrower than that of fixed-interval smoothing.

\section{Gaussian-Mixture Noise Case}

\subsection{The Gaussian-sum filter}

State-space model with non-Gaussian noise distribution such as the Pearson family of distributions 
can provide resonalble estiamtes even with the presence of abrupt stractural changes 
or outlying observation.
Non-Gaussian filter and smoother using numerical integration can provide very accurate filtering and smoothing distributions at least for low-dimensional state-space models (Kitagawa 1987. Kitagawa and Gersch 1996).
However, for higher order state-space models such as the one for the seasonal adjustment
of monthly data, the application of this numerical integration method is
impractical due to the huge amount of computation involved in numerical integration.
One practical way to mitigate this computational burden is the use of a
Gaussian-sum filter (Sorenson and Alspach (1971), Alspach and Sorenson (1972),
Harrison and Stevens (1976) and Anderson and Moore (1979)). In Kitagawa
(1989), it was shown that such a Gaussian-sum filter can be easily derived from
the non-Gaussian filter algorithm by using Gaussian mixture approximations to the related
densities. Specifically, the following Gaussian mixture representation of the noise distributions
and the state distributions were used:
\begin{eqnarray}
  p(v_n) &=& \sum_{i=1}^{K_v} \alpha_i \varphi_i(v_n), \qquad
  p(w_n)  =  \sum_{j=1}^{K_w} \beta_j \varphi_j(w_n), \nonumber \\
  p(x_n|Y_{n-1}) &=& \sum_{k=1}^{L_n} \gamma_{kn} \varphi_k(x_n|Y_{n-1}), \qquad
  p(x_n|Y_{n})    =  \sum_{\ell =1}^{M_n} \delta_{\ell n} \varphi_\ell (x_n|Y_n) .        
\end{eqnarray}
Here $\varphi_i$ denotes a properly defined Gaussian density and $K_v$, $K_w$,
$L_n$ and $M_n$ are numbers of Gaussian components for the distributions of
the system noise, observation noise, predictive distribution and filter
distribution, respectively. 
Substituting these into (2) and (3), 
we obtain the following algorithm for the Gaussian-sum filtering.

\vspace{2mm}
\textbf{Gaussian-sum one step ahead prediction}
\begin{eqnarray}
    p(x_n|Y_{n-1}) &=& \sum_{i=1}^{K_v} \sum_{\ell =1}^{M_{n-1}} \gamma_{i\ell ,n} \varphi_{i\ell}(x_n|Y_{n-1})  \nonumber \\
     &\equiv&  \sum_{k=1}^{L_n} \gamma_{kn} \varphi_k(x_n|Y_{n-1}) , \label{Eq_Gaussian-sum-prediction}
\end{eqnarray}
where $\gamma_{i,\ell ,n} = \alpha_i \delta_{\ell ,n-1}$ and $\varphi_{i\ell}(x_n|Y_{n-1})$ 
is the one-step-ahead predictor of $x_n$
obtained under the assumptions that the filter of $x_{n-l}$ is $\varphi (x_{n-1}|Y_{n-1}) \sim
N(x_{n-1|n-1}^{\ell},V_{n-1|n-1}^{\ell})$ and that $p(v_n)= \varphi_i(v_n) \sim N(0, Q_i)$. 
Therefore, $\varphi_{i\ell }(x_{n-1}|Y_{n-1})$ is also Gaussian and its mean and variance-covariance matrix can be obtained by the ordinary Kalman predictor:
\begin{eqnarray}
  x_{n|n-1}^{i\ell} &=& F_n x_{n-1|n-1}^\ell , \nonumber \\
  V_{n|n-1}^{i\ell} &=& F_n V_{n-1|n-1}^\ell F_n^T + G_n Q_i G_n^T. \label{Eq_GSPred}
\end{eqnarray}
In (\ref{Eq_Gaussian-sum-prediction}), $\equiv$ means re-numbering the double summation by a single summation, and
therefore $L_n = K_v M_{n-1}$.

\vspace{2mm}
\textbf{Gaussian-sum filtering:}
\begin{eqnarray}
   p(x_n|Y_{n}) &\propto&  \sum_{j=1}^{K_w} \sum_{k=1}^{L_n} \delta_{jk,n}\varphi_{jk} (x_n|Y_n) \nonumber \\
                &\equiv&  \sum_{\ell =1}^{M_n} \delta_{\ell n} \varphi_\ell (x_n|Y_n) , 
\end{eqnarray}
where, $\delta_{jk,n} = \beta_j \gamma_{kn} \varphi_{jk} (y_n|Y_{n-1})$, $M_n= K_w L_n$ 
and $\varphi_{jk}(x_n|Y_n) \sim N(x_{n|n}^{jk}, V_{n|n}^{jk})$ 
is the filter of $x_n$ obtained under the assumption that the one-step-ahead predictor
of $x_n$ is $N(x_{n|n-1}^k ,V_{n|n-1}^k)$ and $p(w_n)= \varphi_j(w_n) \sim N(0,\sigma_j^2)$.
Therefore, it can be obtained by the following Kalman filter:
\begin{eqnarray}
K_n^{jk}     &=& V_{nln-1}^k H_n^T(H_nV_{n|n-1}^k H_n^T + \sigma^2_j)^{-1}, \nonumber \\
x_{n|n}^{jk} &=& x_{n|n-1}^k + K_n^{jk}(y_n - H_n x_{n|n-1}^k), \label{Eq_GSFilt}\\
V_{n|n}^{jk} &=& (I-K_n^{jk} H_n)V_{n|n-1}^k. \nonumber
\end{eqnarray}

The advantage of the Gaussian-sum filter is that the parameters of each component can be estimated by the Kalman filter. However, this method has a serious drawback.
Namely, the number of Gaussian components $L_n$ and $M_n$ increases $K_v \times K_w$ times at each time step of filtering.
Thus, the number of Gaussian components increases exponentially with time. A computationally efficient way to reduce the number of Gaussian components is essential.
And a practical measure to deal with this is to reduce the number of components to a pre-determined number $M_{max}$ at each time step.
There is much research on Gaussian component reduction algorithms and the criteria for selecting pairs of Gaussian components to merge (Kitagawa (1994, 2020), Runnalls (2007), Salmond (1990), Williams and Maybeck (2003)).

\subsection{Gaussian-sum smoother}

We now consider the development of a Gaussian-sum smoother. Since the
Gaussian-sum version of the filter has been already given in (\ref{Eq_GSPred}) and (\ref{Eq_GSFilt}), it suffices
to show the implementation of smoothing (4).
Assume that $p(x_n |Y_{n -1})$ and $p(Y_{n:N}|x_n)$ are expressed by
\begin{eqnarray}
   p(x_n|Y_{n-1}) &=& \sum_{k=1}^{L_n} \gamma_{kn} \varphi_k(x_n|Y_{n-1}),\nonumber \\ 
   p(x_n|Y_{n:N})   &=& \sum_{\ell =1}^{M_n} \delta_{\ell n} \varphi_\ell (x_n|Y_n) , 
\end{eqnarray}
where $\varphi_k(x_n|Y_{n-1}) \sim N(x_{n|n-1}^k,V_{n|n-1}^k)$ and $\varphi_\ell (Y_{n:N}|x_n)
\sim N(z_{n|n}^\ell ,U_{n|n}^\ell )$. Then
by analogy to the derivation of the Gaussian-sum filter (\ref{Eq_GSFilt}), for the state-space
model
\begin{eqnarray}
  x_n &=& F_n x_{n-1} + G_nv_n \nonumber \\
  z_n &=& x_n + w_n,
\end{eqnarray}
the Gaussian-sum smoother is obtained by
\begin{eqnarray}
      p(x_n|Y_N) &\propto&  p(Y_{n:N}|x_n) p(x_n|Y_{n-1}) \nonumber \\
         &=& \sum_{\ell =1}^{M_n} \sum_{k=1}^{L_n} \delta_{\ell k}\gamma_{kn} \varphi_k (x_n|Y_{n-1})\varphi_\ell (x_n|Y_{n-1:N}) \nonumber \\
         &=&  \sum_{\ell =1}^{M_n}\sum_{k=1}^{L_n}  \delta_{\ell n}\gamma_{\ell k} \varphi_{\ell k} (x_n|Y_N) . 
\end{eqnarray}
Here $\varphi_{\ell k} (x_n|Y_N)$ is the Gaussian density whose mean and the variance-covariance matrix are
obtained by replacing $H_n$ by $I$ and $y_n$ by $z_{n|n}^\ell$,
\begin{eqnarray}
J_n^{\ell k} &=& V_{n|n-1}^k (V_{n|n-1}^k + U_{n|n}^\ell )^{-1} \nonumber \\
x_{n|N}^{\ell k} &=& x_{n|n-1}^k + J_n^{\ell k}(z_{n|n}^\ell - x_{n|n-1}^k) \\
V_{n|N}^{\ell k} &=& (I-J_n^{\ell k}) V_{n|n-1}^k.
\end{eqnarray}

\subsection{Examples of Gaussian-sum smoother}


\vspace{2mm}
We next consider the behavior of the Gaussian-sum smoother when there exist
some jumps of the trend in seasonal data. 
For that purpose, we consider the BLSALLFOOD data with artificially generated two jumps, i.e.,
\begin{eqnarray}
   \bar{y}_n = \left\{ \begin{array}{ll}
              y_n       &\quad  n=1,\ldots ,79 \\
              y_n + 150 &\quad  n=80,\ldots ,100 \\
              y_n - 100 &\quad  n=101,\ldots , N. \end{array}   
        \right.
\end{eqnarray}

The system noise for the trend component and the observation noise are assumed to follow
\begin{eqnarray}
    u_n \sim \sum_{j=1}^{m_q} \alpha_j N(0,\tau^2_{1j}), \qquad
    w_n \sim \sum_{\ell=1}^{m_r} \beta_{j\ell} N(0,\sigma^2_{\ell}) ,
\end{eqnarray}
where $\alpha_1 + \alpha_2 = 1$ and $\beta_1 + \beta_2 = 1$.
If $m_Q$ or $m_R$ is 1, the noise distribution becomes the ordinary Gaussian distribution.
If it is set to 2, the noise distriution becomes a mixture of two Gaussian distirbutions,
with ordinary variance and a big variance corresponding to occasional jump of the trend
or outliers of the observations.
 The variances $\tau_{j1}^2, \,(j=1,2,3)$ and $\sigma_{1}^2$ are estimated by the
maximum likelihood method, but for simplicity, the large variances $\tau^2_{12}$ and
$\sigma^2_2$ are arbitralily set to large values.

\begin{figure}[tbp]
\begin{center}
\includegraphics[width=160mm,angle=0,clip=]{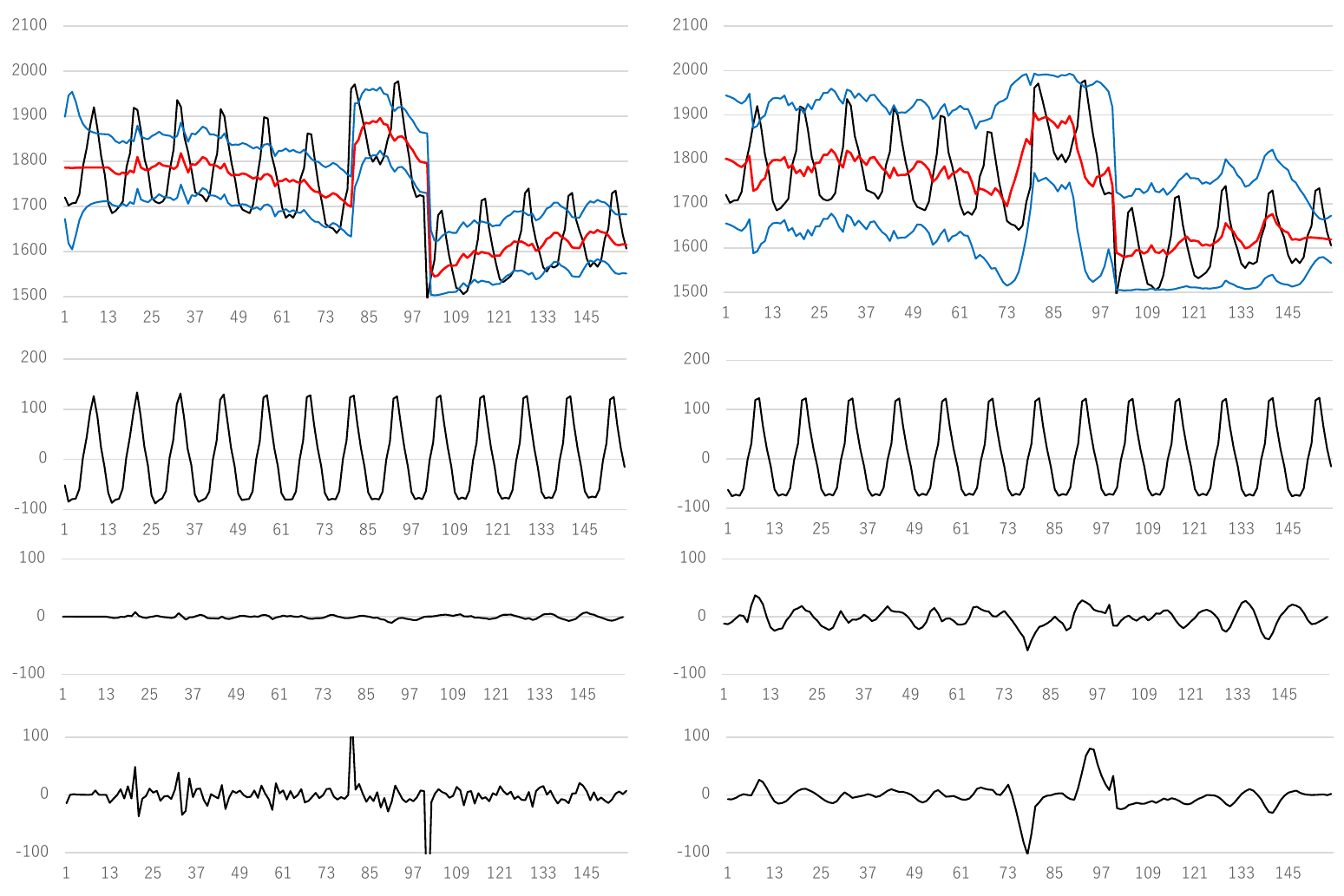}
\end{center}
\caption{Estimates by the forward predictor and the backward filter
for the Gaussian-mixture model for seasonal adjustment with $m_q=2$ and $m_r=1$.
Left plots: forward predictor, Right plots: backward filter. 
Top plot shows the data (black), the mean of the trend (red), $\pm 2$ standard error (blue),
the second plot the seasonal component, the third plot the AR component and the bottom plot shows the noise component. }
\label{Fig_Gaussian-sum_forward_backward_filters}
\end{figure}

\begin{table}
\caption{Estimated parameters of three seasonal adjustment models}
\label{Tab_Estimated-parameters}
\begin{center}
\begin{tabular}{c|c|c|c}
              & $m_q=1$, $m_r=1$        & $m_q=2$, $m_r=1$       & $m_q=2$, $m_r=2$      \\
\hline
$\tau_{11}^2$ &     99.520              &     0.32124            &     0.21809           \\
$\tau_{12}^2$ & \rule[1mm]{8mm}{0.05mm} & $1.0 \times 10^5$      & $1.0 \times 10^5$     \\
$\tau_{21}^2$ & $0.94276\!\times\! 10^{-6}$ & $0.94276\!\times\! 10^{-6}$& $0.12253\!\times\! 10^{-3}$\\
$\tau_{31}^2$ &     43.030              &     43.030             &     43.030             \\
$\sigma_1^2$  &     23.770              &     15.916             &     15.636             \\
$\sigma_2^2$  & \rule[1mm]{8mm}{0.05mm} & \rule[1mm]{8mm}{0.05mm}& $1.0 \times 10^5$      \\
$a_1$         &    1.14852              &     1.17769            &     1.14850            \\
$a_2$         &  $-0.33418$             &    $-0.33438$          &    $-0.33415$          \\
\hline
\end{tabular} 
\end{center}
\end{table}

Table \ref{Tab_Estimated-parameters} shows the estimated parameters of three
seasonal adjustment models.
Figure \ref{Fig_Gaussian-sum_forward_backward_filters} shows the estimates of trend,
seasonal component, AR component and  the noise obtained by the forward predictor (left plots) and
the backward filter (right plots) when we assumed the seasonal adjustment model with second order
AR model. 
Even the forward predictor can adapt to the level shift of the trend.
However, there is a delay of response at least one time point and 
$\pm$2 confidence interval is vary large.
Further, almost no  AR component is detected.
The backward filter is very variable and trend estimaate has very larde $\pm$2 confidence interval. 
This may be due to the fact that the AR component model of the backward-looking stat--space model is non-stationary and contains divergent component.

\begin{figure}[tbp]
\begin{center}
\includegraphics[width=160mm,angle=0,clip=]{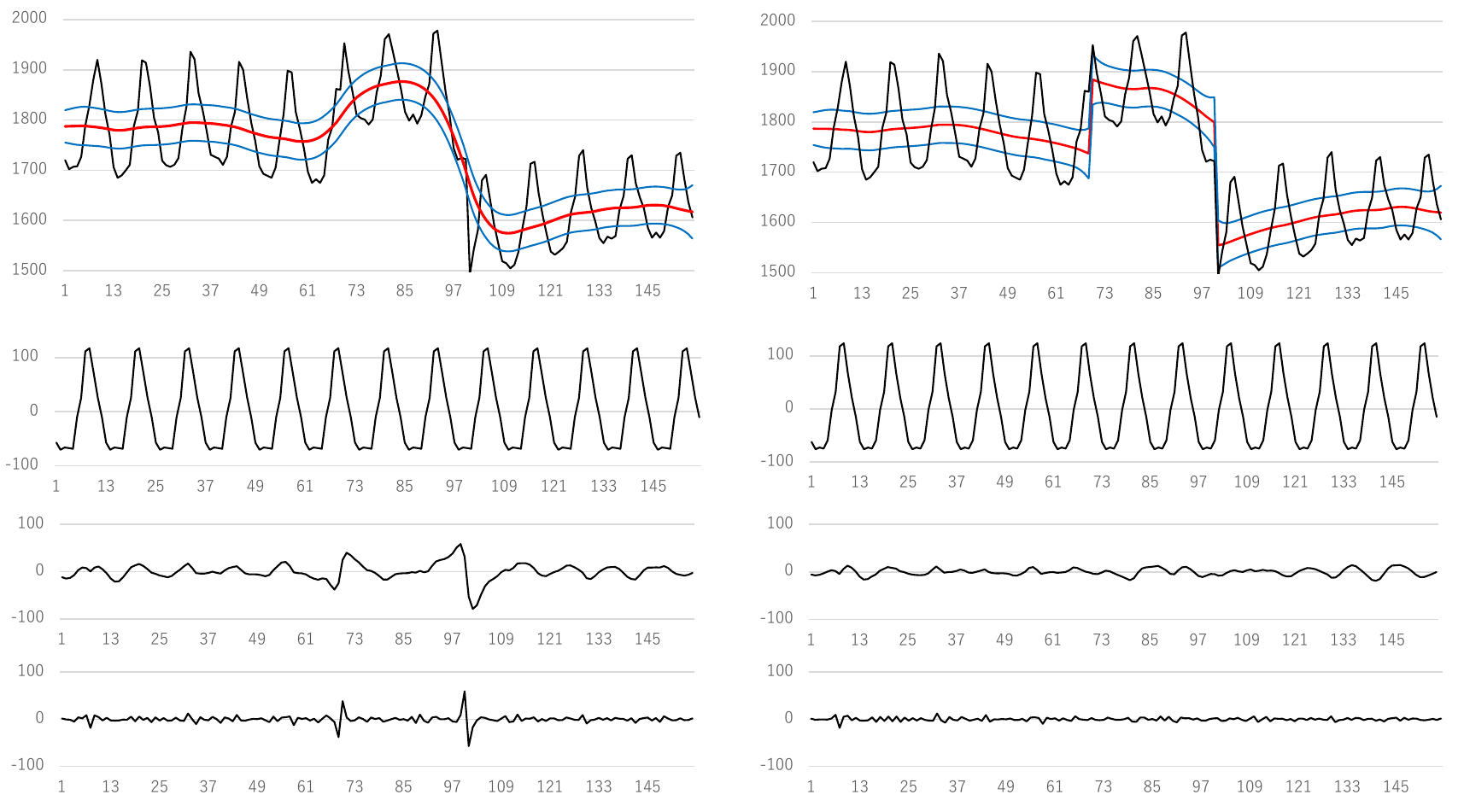}
\end{center}
\caption{The seasonal adjustment by the two-filter formula for smoothing
for the Gaussian-mixture model for seasonal adjustment.
Left plots: Gaussian noise model ($K_v=1$, $K_w=1$), Right plots: Gaussian-mixture noise model for system noise ($K_v=2$, $K_w=1$), $M_{max}=6$. 
Top plot shows the data (black), the mean of the trend (red), $\pm 2$ standard error (blue), 
the second plot the seasonal component, the third plot the AR component and the bottom plot shows the noise component. 
Bottom left plot: Gaussian-sum smoothing with $K_v=2$, $K_w=1$, $M_{max}=1$, 
Bottom right plot: Gaussian-sum smoothing with $K_v=2$, $K_w=1$, $M_{max}=2$.
}
\label{Fig_Gaussian-sum_two-filter-formula}
\end{figure}

Left plots of  show the results obtained by
the two-filter smoothing for the model with $m_Q=1$ and $m_R=1$, i.e.,
ordinary Gaussian model. 
The estimated trend is very smooth, but instead does not respond to rapid changes of the mean of the time series.
Thus, the AR and noise components show unnatural spikes at the time of structural change.
On the other hand, right plots show the results
by the model with $m_Q=2$ and $m_R=1$, i.e.,
a Gaussian mixture distribution for the trend component and ordinary Gaussian distribution for the observation noise.
With this Gaussian-mixture modeling. the estimated trend is very smooth, yet the two jumps in the trend are clearly detected by this model.
As a result, the AR and noise components are extremely natural.

\begin{figure}[tbp]
\begin{center}
\includegraphics[width=160mm,angle=0,clip=]{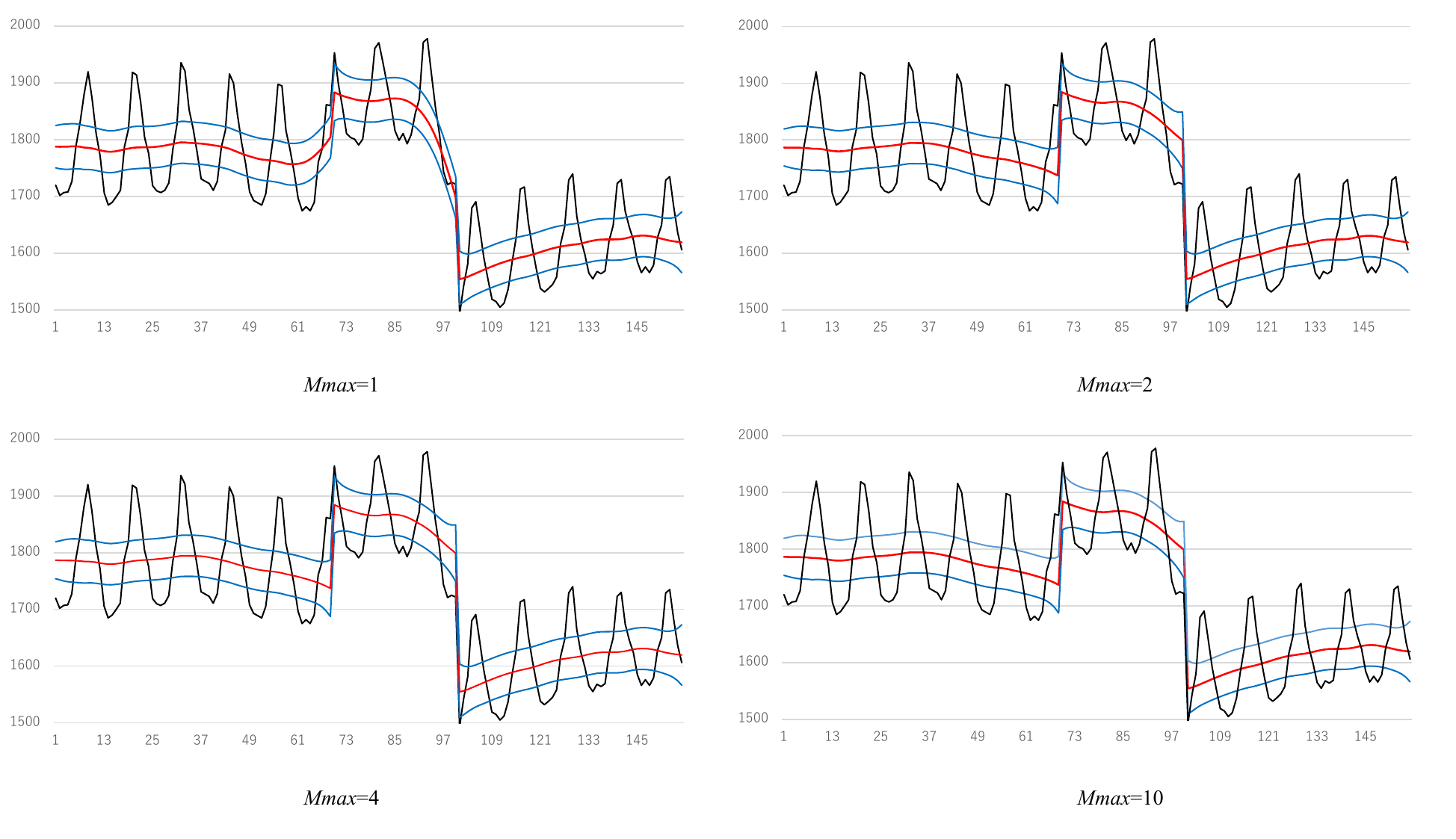}
\end{center}
\caption{Effect of the number of Gaussian mixture components, $M_{max}$.
Left plot: Gaussian-sum smoothing with $K_v=2$, $K_w=1$, $M_{max}=1$, 
Right plot: Gaussian-sum smoothing with $K_v=2$, $K_w=1$, $M_{max}=2$.
}
\label{Fig_Gaussian-sum_two-filter-formula2}
\end{figure}

Figure \ref{Fig_Gaussian-sum_two-filter-formula2} shows the change in decomposition due to Gaussian-sum smoothing as the number of Gaussian components, $M_{max}$, is varied for the model with $K_v=2$ and $K_w=1$.
The upper left plot is the case where $M_{max}=1$, i.e., actually approximated by a single Gaussian distribution.
Unlike the usual Gaussian model, changes in trend can be detected, but they are not clear jumps.
In contrast, the upper right, lower left, and lower right plots in Figure 7 show the cases of $M_{max}$=2, 4, and 10, respectively. Two jumps of the trend component are clearly detected when $M{max}$=2, and the results hardly change when $M_{max}$=4, 10.
This means that, at least in this example, $M_{max}$=2 is sufficient.

\section{Particle Filter and Smoother}
In this section, we consider particle filters applicable to state estimation for a wide range of state-space models (Gordon et.al.(1993), Kitagawa(1996)). However, for the sake of simplicity, we will consider linear non-Gaussian state-space model here:
\begin{eqnarray}
  x_n &=& F_n x_{n-1} + G_n v_n \nonumber \\
  y_n &=& H_n x_n + w_n,
\end{eqnarray}
where $v_n \sim q(v)$ and $w_n \sim r(w)$.

\subsection{Particle filter for state-space model with non-Gaussian noise distributions}

In contrast to other nonlinear or non-Gaussian filters, 
the particle filter approximates each density function by
a large number of particles that can be considered as realizations from
the true distribution. 
Assume that $m$ is the number of particles, $k$ is the state dimension,
$\ell$ is the dimension of the system noise and $N$ is the number of observations.
Specifically, we approximate each of the predictive distribution, the filter distribution,
smoothing distribution and the system noise distribution by $m$ particles as 
$\{p_k^{(1)},\ldots , p_k^{(m)}\} \sim p(x_{k}|Y_{k-1})$, 
$\{f_k^{(1)},\ldots , f_k^{(m)}\} \sim p(x_{k}|Y_{k})$,  
$\{s_{k|N}^{(1)},\ldots , s_{k|N}^{(m)}\} \sim p(x_{k}|Y_{N})$ and 
$\{v_k^{(1)},\ldots, v_k^{(m)}\} \sim q(v_{n})$.
The particle filter for the linear non-Gaussian state-space model is given as follows:
\begin{enumerate}
\item {\em Generate $k$-dimensional random number that approximates initial state
density, $f_0^{(j)} \sim p_0(x)$, for $j=1,\ldots ,m$.}
\item {\em Repeat the following steps for $n=1,\ldots ,N$};
  \begin{enumerate}
  \item [2-1] {\em Generate a $k$-dimensional random number that approximates the system noise $v_n^{(j)} \sim q(v)$, for $j=1,\ldots ,m$}.
  \item [2-2] {\em Compute $p_n^{(j)} = F_n f_{n-1}^{(j)} + G_n v_n^{(j)}$ to obtain a particle approximating the predictive distribtion $p(x_n|Y_{n-1})$, for $j=1,\ldots ,m$.}
  \item [2-3] {\em Compute the importance weight $\alpha_n^{(j)} = r(y_n- H_np_n^{(j)})$,
        for $j=1,\ldots ,m$.}
  \item [2-4] {\em Generate $f_n^{(j)}$, $j=1,\ldots ,m$
by the resampling of $p_n^{(1)},\ldots ,p_n^{(m)}$},
{\em namely, for $j=1,\ldots ,m$, repeat the following steps}:
  \begin{enumerate}
  \item [R-1] {\em Generate uniform random number,} $u_n^{(j)} \in U[0,1)$.
  \item [R-2] {\em Search for the integer i that satisfies} 
  $\displaystyle \sum_{\ell =1}^{i-1} \tilde\alpha_n^{(\ell )} < u_n^{(j)}
  \leq \sum_{\ell =1}^{i} \tilde\alpha_n^{(\ell )}$.
   \item [R-3] {\em Obtain a particle approximating the filter by setting} $f_n^{(j)} = p_n^{(i)}$. 
  \end{enumerate}
  \end{enumerate}
\end{enumerate}
In step R-2, $\tilde\alpha_n^{(\ell )}$ is the normalized $\alpha_n^{(j)}$ defined by
$\tilde\alpha_n^{(\ell )} = \displaystyle \frac{1}{m}\sum_{j=1}^m \alpha_n^{(j)}$.

\subsection{Fixed-lag particle smoother}

In this subsection, we briefly present fixed-lag smoothing algorithm
that is obtained by a simple modification of the particle filter
algorithm (Kitagawa (1996)).
Assmue that $S_{i:n}^{(j)} \equiv (s_{i|n}^{(j)}, \cdots, s_{n|n}^{(j)})$ 
denotes the $j$-th particle of the conditional joint distribution 
$p(x_i, \cdots, x_n|Y_n)$. 
Then, an algorithm for fixed-lag smoothing is obtained by 
replacing step (2-4) of the particle filter algorithm with

\vspace{2mm}

\noindent\hangindent=11mm
(2-4L) {\em Generate $\{ S_{n-L:n}^{(j)}, j=1,\ldots ,m\}$ 
by resampling $\{(S_{n-L:n-1}^{(j)},p_n^{(j)}), j=1,\ldots ,m\}$
using the importance weights $\{\alpha_n^{(j)},$ $j=1,\ldots ,m\}$.}

\vspace{2mm}

\noindent
Note that in this modification, 
it is necessary to store the past $L$ states $S_{n-L:n}^{(j)}, j=1,\ldots ,m$,
but we can use the same importance weight $\alpha_n$ 
as that used in step (2-4).
Note also that to obtain the smoothing distribution, it is not necessary to store all components of the $m$-dimensional particle $S_n^{(j)}$, but only those components that we want to estimate directly, such as the trend and seasonal components.

If we set $L=n$ and use entire particles, this fixed-lag smooting is
identical to the fixed-interval smoothing.
However, this causes two difficulties in the implementation of the smoother.
Firstly, we need a large memory to store entire particles 
generated in the filtering process.
Secondly, and more importantly, since the number of particles, $m$, is finite,
by repeating the resampling step (2-4L), the number of different
particles in $\{S_{n}^{(1)},\ldots ,S_{n}^{(m)}\}$ gradually decreases
monotonically, eventually resulting in degradation of the accuracy of the distribution (Kitagawa (1996)).
The effect of the particle size, $m$, on this phenomenon is reported
in details in Kitagawa (2014).

\subsection{The two-filter formula for smoothing}

As mentioned in the previous section, the main reason that the smoothing distribution 
loses its accuracy is the collapse of distribution, 
namely, the reduction of the number of different particles
as the result of the repeated resampling step.
Various resampling algorithms have been developed to mitigate this difficulty
(Doucet et.al. (2001)).
One way to address this problem is to use two-filter formula based on the
decomposition of $ p(x_n|Y_N)$ (Fraser (1967), Kitagawa (1994, 1996, 2014), Balenzuela et.al. (2022)): 
\begin{eqnarray}
 p(x_n|Y_N) \propto p(x_n|Y_{n-1})p(Y_{n:N}|x_n) ,
\end{eqnarray}
where \( Y_{n:N} \equiv \{y_{n},...,y_N\}$. 
Here $ p(Y_{n:N}|x_n) \) can be computed by filtering backward in time.

To precisely obtain smoothed posterior distribution $p(x_n|Y_N)$,
it is necessary to compute the importance weight of each particle of the 
predictive distribution $p_n^{(j)}$ by
\begin{eqnarray}
\beta_n^{(j)} = \frac{1}{m}\sum_{i=1}^m D( \tilde f_n^{(i)} - p_n^{(j)}),
\label{Eq_ParticleSmoother_beta}
\end{eqnarray}
where $\tilde f_n^{(i)}$ is a particle generated by backward filtering
and $D(e)$ is defined by $D(e) = q_n(G_n^Te)$.
For example, for the case of seasonal adjustment model with AR component,
\begin{eqnarray}
q_n(G_n^T (\tilde f_n^{(i)} - p_n^{(j)}) 
  \sim  N(e_n^{t},\tau^2_1) N(e_n^{s},\tau^2_2) N(e_n^{p},\tau^2_3) ,
\label{Eq_ParticleSmoother_beta2}
\end{eqnarray}
where $e_n^{t}$, $e_n^{s}$ and $e_n^{p}$ are the first, the third and
the 14-th element of $\tilde f_n^{(i)} - p_n^{(j)}$, respectively.
The evaluation of the likelihood for all $m^2$ pairs of 
particles requires a huge amount of computations for large $m$. 

In practice, however, it can be approximated reasonably, by sampling
$r$ particles from $m$ particles, $\tilde f_n^{(1)}, \ldots,
\tilde f_n^{(m)}$,
\begin{eqnarray}
{\beta^\prime}_n^{(j)} = \frac{1}{r}\sum_{\alpha =1}^r D( \tilde f_n^{(i_\alpha)} - p_n^{(j)}).
\end{eqnarray}
It is possible to achieve good approximation of the
exact smoothing distribution by computing the importance weights $\beta_{n}^{(j)}$ for only $r=10$ to $10^2$ particles
of $p(Y_{n:N}|x_n)$ (Kitagawa (2014)).

\subsection{Examples of Particle Smooting}

\subsubsection{Trend estimation}
We consider two examples of smoothing problem, a simple trend estimation
in the presence of several jumps and decomposition of seasonal time series 
into the trend, seasonal, stationary AR and the noise components.
The dimensions $m$ of the state vector for these models are 1 and 15, respectively.

In the first example, the test data was generated by the model
\begin{eqnarray}
  y_n = \phi_n + w_n, \quad w_n \sim N(0,1),
\end{eqnarray}
where the mean value function is defined by $\phi_n = 0$ for $1 \leq n \leq 100$, $-1$ for 
$101 \leq n \leq 250$, 1 for $251 \leq n \leq 350$ and 0 for $351 \leq n \leq 500$
(Kitagawa (1996)).

\begin{figure}[tbh]
\begin{center}
\includegraphics[width=140mm,angle=0,clip=]{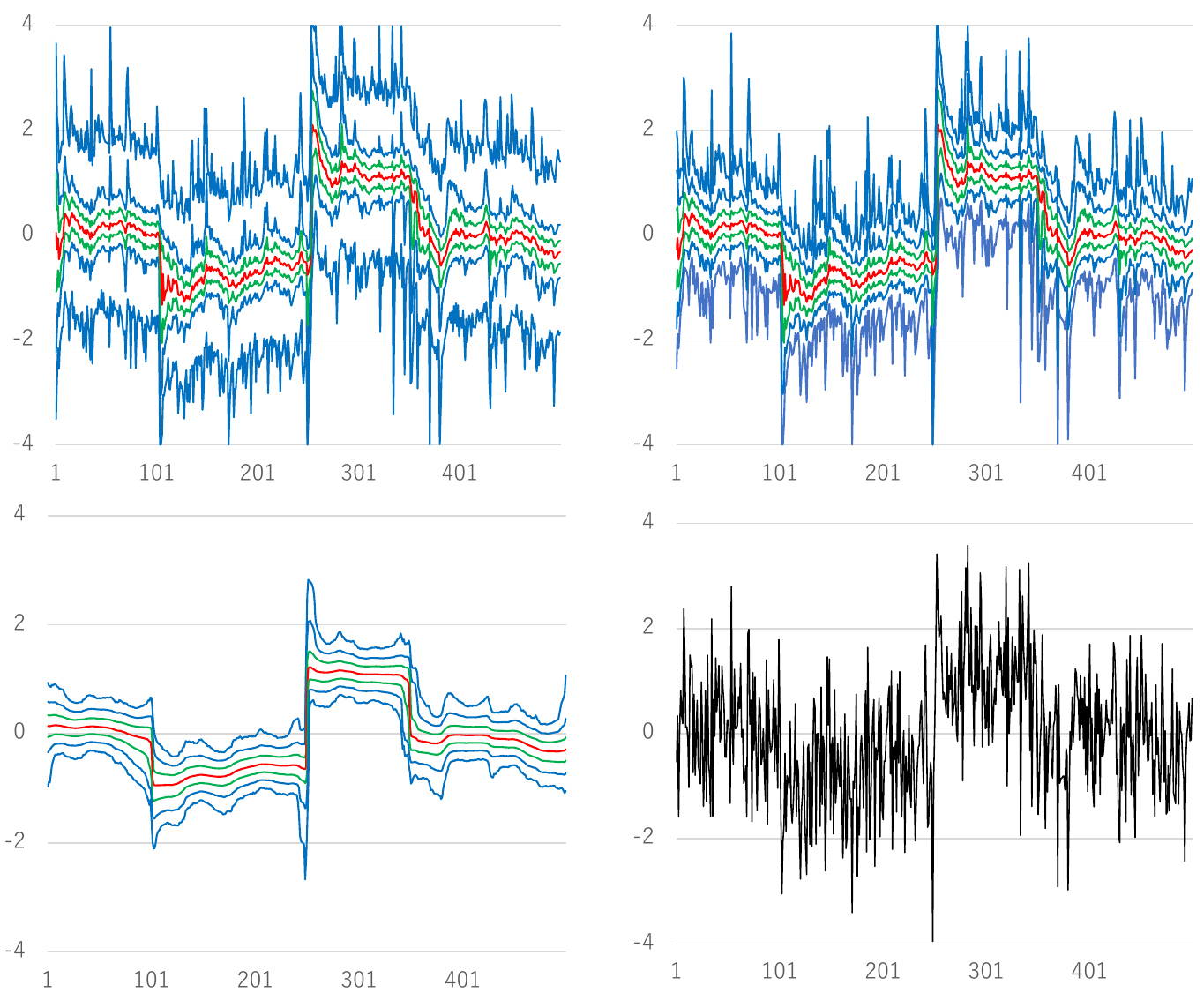}
\end{center}
\caption{Posterior distributions of the trend by the forward predictor, 
the backward filter and 
the two-filter smoother for the trend model of order 1 with Cauchy distribution noise.
Upper left plot: forward predictor, Upper right plot: backward filter, Lower left plot:
two-filter smoother. Lower right plot: test data. }
\label{Fig_Particle_smoother_Trend_model_new}
\end{figure}

Figure \ref{Fig_Particle_smoother_Trend_model_new} shows the test data and the posterior
distributions of the trend obtained by the forward predictor, backward filter and 
the two-filter smoother with the following linear state-space model with
Cauchy distribution system noise:
\begin{eqnarray}
  x_n &=& x_{n-1} + v_n \nonumber \\
  y_n &=& x_n + w_n,
\end{eqnarray}
with $w_n \sim N(0,1.022)$, $v_n \sim Cauchy(0,3.48\times 10^{-1})$ and 
$x_{0|0} \sim N(\hat{\mu},\hat{V})$, where $\hat{\mu}$ and $\hat{V}$ are
the mean and the variance of the test data.
In each plot, the $50\%$ points of the posterior distributions (red) 
and the 0.13, 2.27, 15.87, 84.13, 97.73 and 99.87$\%$ points (blue or green) that correspond to 
$-3, -2, -1, 1, 2, 3$ standard error points for Gaussian distribution
are shown.
The forward predictive distribution is very wiggly and especially the $\pm$3 standard error interval is very wide indicating that the distribution is heavy tailed.
On the other hand, the particle filter with Cauchy distribution noise can adapt to abrupt changes in the mean value function.
The upper right plot shows the results from the backward filter.
The curve obtained is generally similar to the forward predictive distribution,
However, the three standard error intervals are narrower than the forward predictive distribution.

The lower left plot shows the smoothed distribution of the trend component obtained by the two filter smoothing. The smoothed distribution of the trend component obtained by the two filter smoothing algorithms is shown.
Compared with the predictive and filter densities, a very smooth curve is obtained and abrupt changes in the trend component at three points were clearly detected at $n$=101, 256, and 351.


\subsubsection{Seasonal adjustment}

\begin{figure}[tbp]
\begin{center}
\includegraphics[width=160mm,angle=0,clip=]{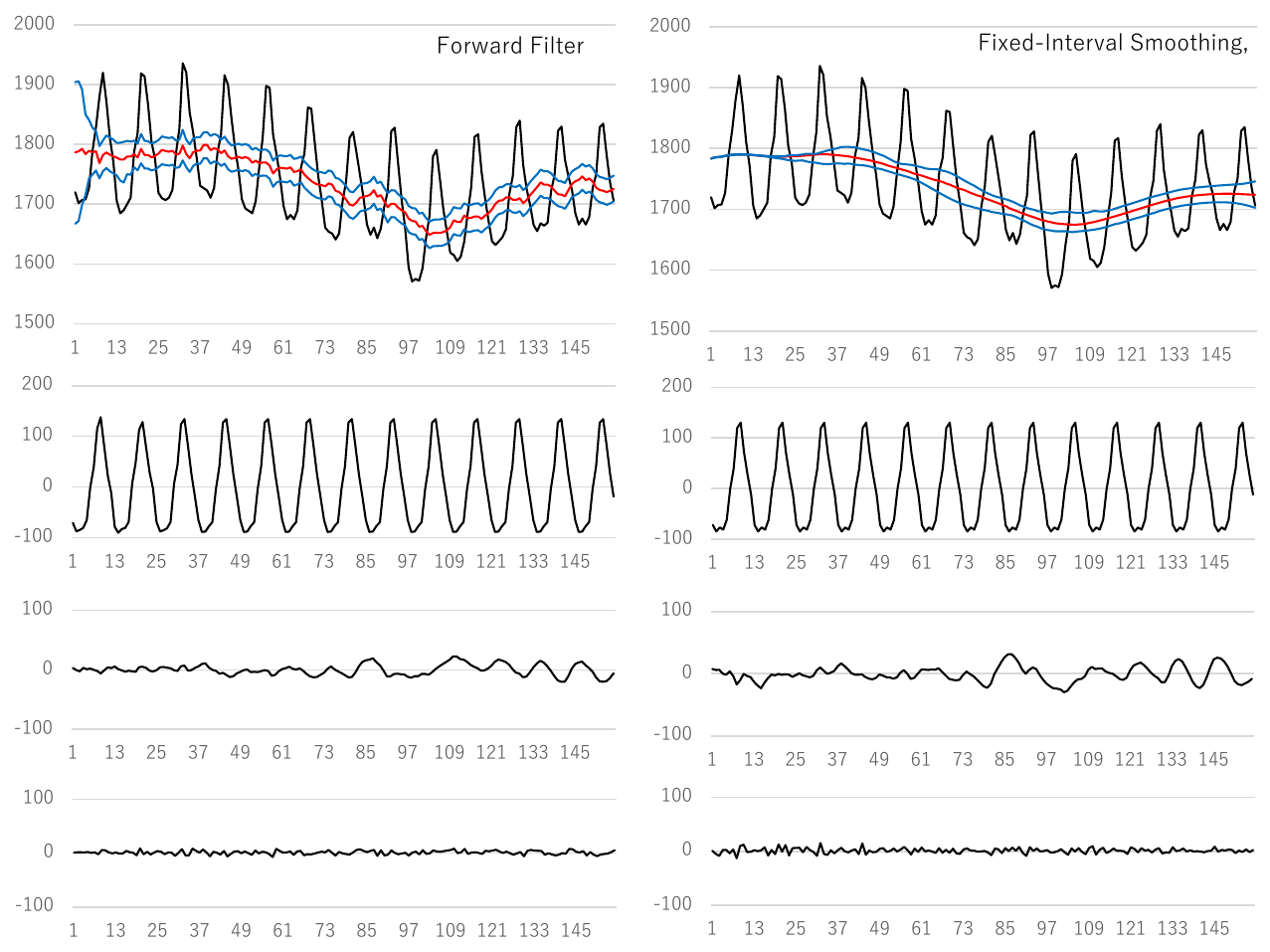}
\end{center}
\caption{The seasonal adjustment by the particle smoother by two-filter formula.
Left plot: without AR component $m_3=0$, Right plot: with AR component, $m_3=2$. 
Top plot shows the data (black), the mean of the trend (red), $\pm 2$ standard error (blue), 
the second plot the seasonal component, the third plot the AR component and the bottom plot shows the noise component. }
\label{Fig_Particle_Filter_Fixed-interval_smoother}
\end{figure}

Figure \ref{Fig_Particle_Filter_Fixed-interval_smoother} shows the decomposition 
of a time series by a seasonal adjustment model with AR component 
($m_1=2$, $m_2=11$ and $m_3=2$) obtained by the particle filter.
Note that in this case, since the model is linear and the noise distributions are
Gaussian, we can obtain the exact posterior distribution without using
particle smoother as shown in Section 3. The purpose of using this model
for particle smoothing is to compare with the exact results obtained by the 
Kalman filter and the smoother.

The left plots show the results by the forward filter and the right plots the
results by the fixed interval smoother.
In either case, from top to the bottom, data and the trend,
seasonal component, AR component and the noise component are shown.
In the case of trend estimates, not only the posterior mean (red) but also
the $\pm$2-standard error interval is shown in blue.
The trend estimate by the forward filter has large fluctuation.
On the other hand, the trend estimate by the fixed-interval smoother
is very smooth and is similar to the one obtained by the Kalman smoother
shown in Figure 3.
However, for $n <24$, the $\pm$2-standard error interval shrinks, indicating that the posterior distribution has collapsed to single particle or very few particles.
This is a typical phenomenon that often occurs in particle smoothing.

\begin{figure}[tbp]
\begin{center}
\includegraphics[width=160mm,angle=0,clip=]{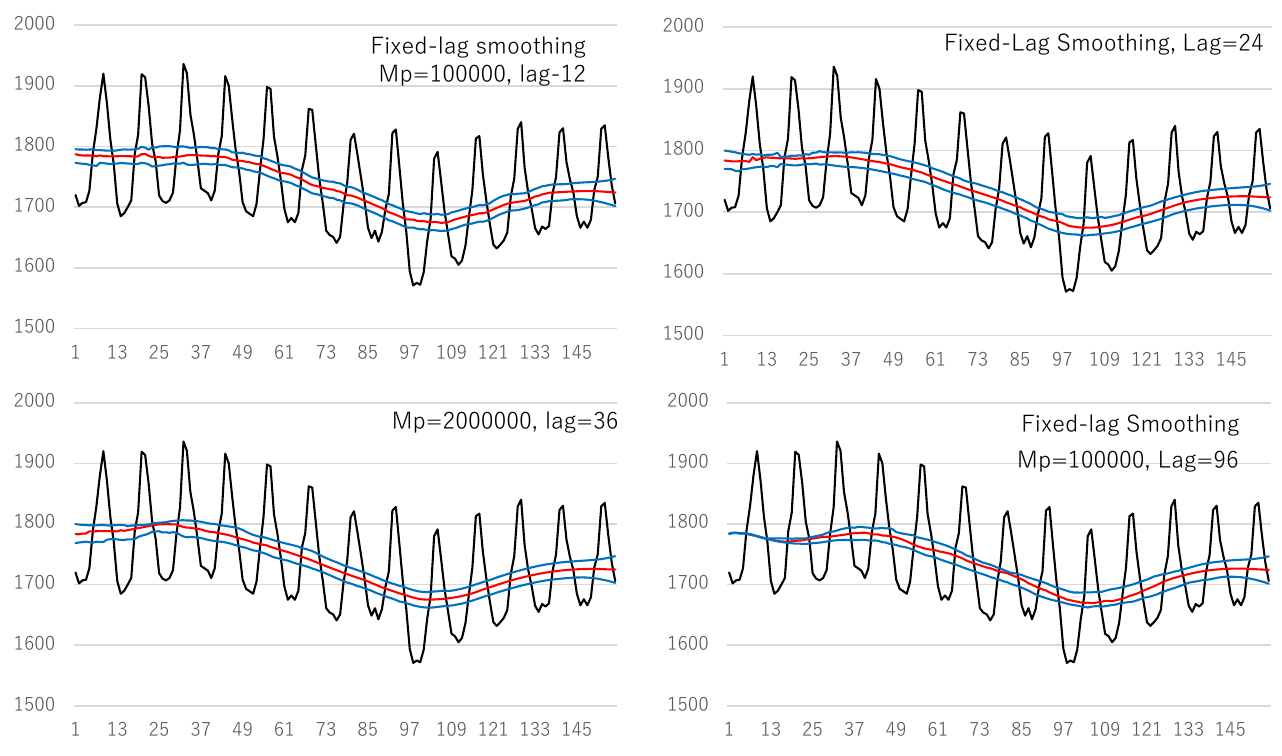}
\end{center}
\caption{The seasonal adjustment by the fixed-lag smoother with lag=12 (upper left), 24 (upper right), 36 (lower left) and 96 (lower right).
Only the data (black) and the mean of the trend (red) and $\pm 2$ standard error (blue) are shown.}
\label{Fig_Fixed-lag_smoother_Trend}
\end{figure}

Figure \ref{Fig_Fixed-lag_smoother_Trend} shows the trend estimates obtained
by the fixed-lag smoothing with four lag length, lag=12, 24, 36 and 96.
The seasonal component, AR components and the noise component are ommited here
because visually there is no significant difference.
From upper left to the lower right, the results obtained by setting lag=12, 24, 36 and 96 are shown. As shown in the upper left plot, even with a small lag such as 12, reasonably smooth estimates are obtained, although it is slightly variable.
When lag is set to 24 or 36, the trend is very smooth, but the posterior distribution of the trend looks degenerate in the first part of the data. 
If the lag is further increased to 96, the posterior distribution degenerates to a single point near the left edge of the data, as is the case with fixed-interval smoothing.
Since fixed lag smoothing estimates the state $x_n$ at time $n$ using data $y_1, \ldots ,y_{n+\mbox{lag}}$, the distribution is likely to degenerate at the left edge of the data.


\begin{figure}[tbp]
\begin{center}
\includegraphics[width=165mm,angle=0,clip=]{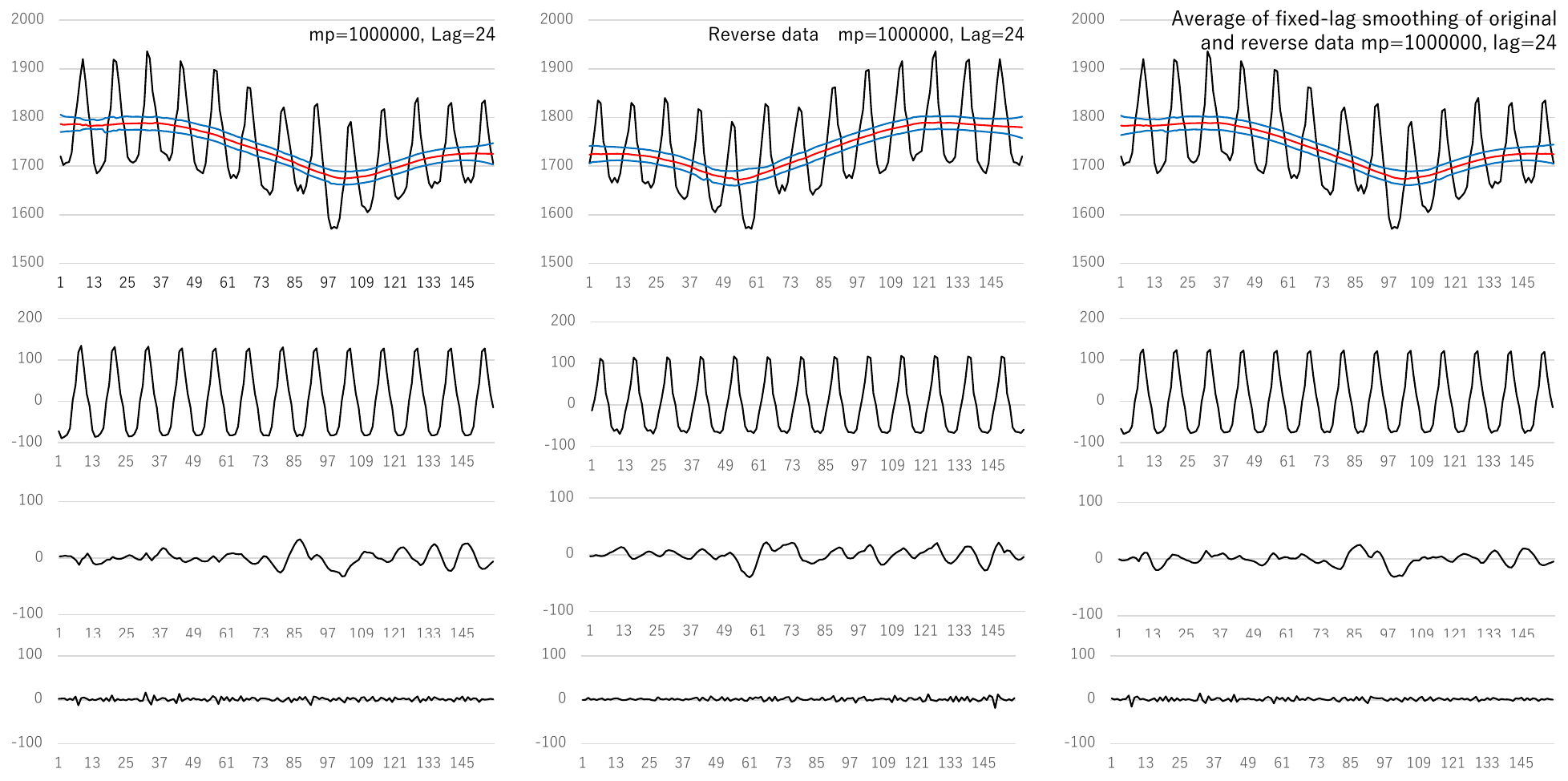}
\end{center}
\caption{The seasonal adjustment by the particle fixed-lag smoothing with 1,000,000 particles and Lag=24.
Left plots: for original time series, Middle plots: for reverse data, Right plots: average of two estimates. 
Top plot shows the data (black), the mean of the trend (red), $\pm 2$ standard error (blue), 
the second plot the seasonal component, the third plot the AR component and the bottom plot shows the noise component. }
\label{Fig_Fixed-lag_average_smoother}
\end{figure}

Figure \ref{Fig_Fixed-lag_average_smoother} shows a simple method of mitigating the
problem of degeneration of particles associated with the fixed-lag smoothing.
The left figure shows the results of fixed-interval smoothing when the number of particles is 1,000,000 and lag=24. Even with a large number of particles, good estimates are not obtained near the left end. The middle figure shows the result of applying fixed-interval smoothing in the same way to time-reversed data, $y_N,\ldots,y_1$.
The right figure is the result of simply averaging the corresponding percentile points obtained by forward and backward fixed-lag smoothing. It can be seen that the estimates are very smooth over the entire interval, similar to the results obtained by Kalman smoothing.

\begin{figure}[tbp]
\begin{center}
\includegraphics[width=165mm,angle=0,clip=]{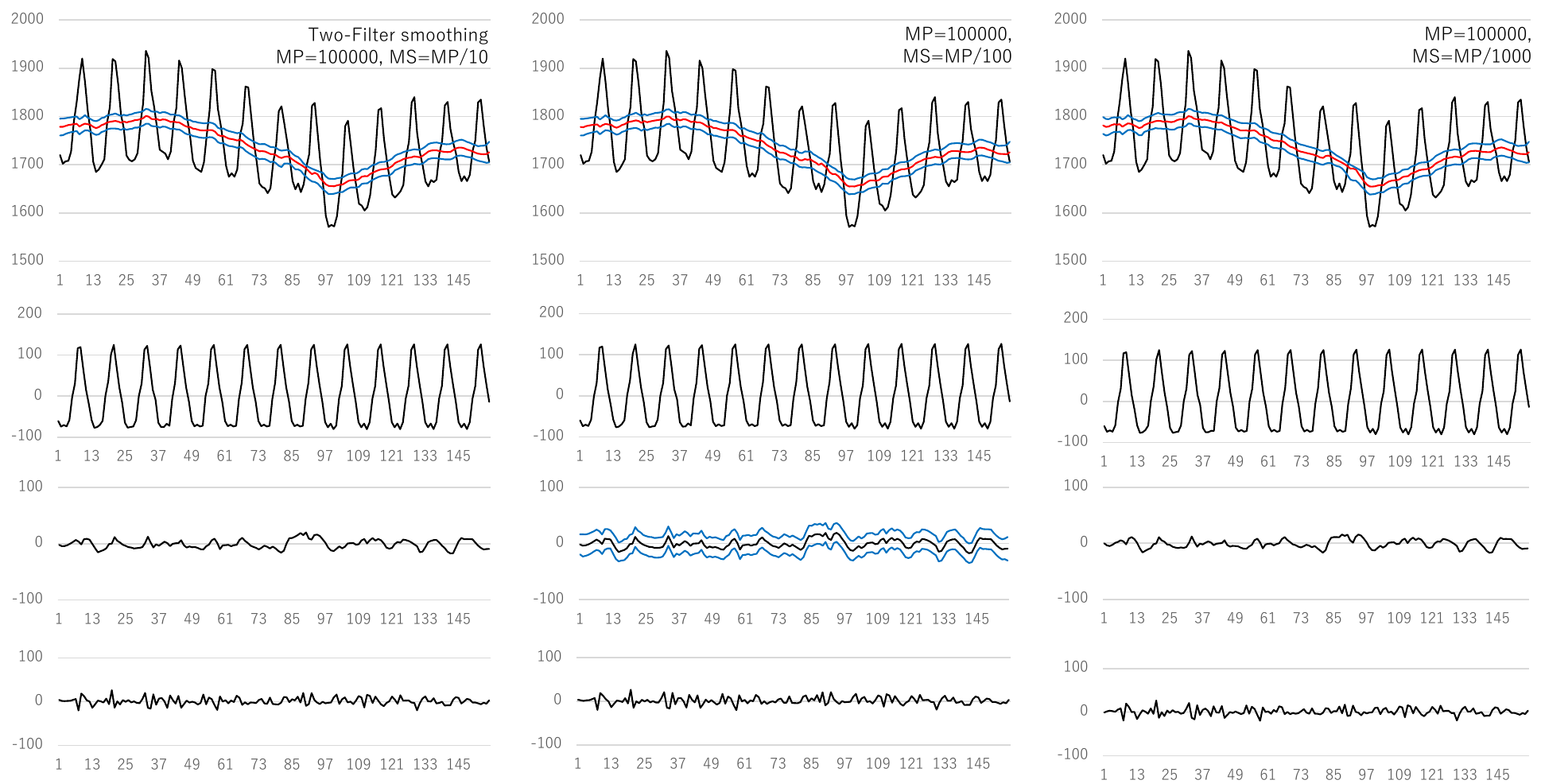}
\end{center}
\caption{The seasonal adjustment by the particle smoother by two-filter formula.
Left plot: $ms=10$, Middle plot: $ms=100$, Right plot: $ms=1000$. 
Top plot shows the data (black), the mean of the trend (red), $\pm 2$ standard error (blue), the second plot the seasonal component, the third plot the AR component and the bottom plot shows the noise component. }
\label{Fig_Two-Filter_smoother}
\end{figure}

Figure \ref{Fig_Two-Filter_smoother} shows the results obtained by the two-filter
formula for smoothing.
Three different values of $r$ (the number of particles used to evaluate importance weights of the
paricles for forward predictor), 10, 100 and 1000 are shown.
It can be seen that small number of $r$ is sufficient to compute the two-filter 
smoother. However, the posterior distribution
obtained by this method is considerably variable, compared with those obtained by
fixed-lag smoother.

This may be due to the fact that when calculating the importance weight (\ref{Eq_ParticleSmoother_beta2}) for each particle, there is a high probability that either $N(e_n^{t},\tau^2_1)$, $N(e_n^{s},\tau^2_2)$ or $N(e_n^{p},\tau^2_3)$ will become zero for the majority of particles, which is likely to cause degeneracy in the distribution. Further research is needed to solve this problem.

\section{Conclusion}
In this paper, we revisited smoothing algorithms for state-space models, focusing on two-filter formulas, using real examples.
In the case of linear and Gaussian state-space models, fixed-interval smoothing can be performed without using the two-filter formula, but it was confirmed that similar posterior distributions can be obtained by appropriately defining an inverse filter.
In the case of linear and non-Gaussian state-space models, we showed that Gaussian sum smoothing can be achieved even for relatively high dimensional state-space model by setting the inverse filter appropriately.
The two-filter formula is also applicable to the case of particle filters, which can be applied to nonlinear state-space models, but so far better results than smoothing with the two-filter formula have been obtained by fixed-interval smoothing or by averaging forward and backward fixed-lag smoothing distributions.

\vspace{15mm}\newpage
\noindent{\Large\bf Aknowledgements}

This work was supported in part by JSPS KAKENHI Grant Number 18H03210.
Part of this study was conducted during the author was affiliated with Mathematics and Inofrmatics Center at the University of Tokyo.

\end{document}